\documentclass[a4paper,11pt]{article}
\usepackage{paralist, url, comment}
\usepackage{aaskaiid}
\usepackage{threeparttable}
\usepackage{orcidlink}

\newcommand{\kms}{km\,s$^{-1}$}
\newcommand{\meth}{$\mathrm{CH_3OH}$}
\newcommand{\form}{$\mathrm{H_2CO}$}

\newcommand{\blue}{\color{blue}}

\title{Cosmic Rulers: Masers as Tools for Probing Structure in the Galaxy and Beyond, from AU to kpc}
\ShortTitle{Cosmic rulers: Masers as Tools for Probing Structure in the Galaxy and Beyond}

\author[1]{Kazi L.~J.~Rygl\orcidlink{0000-0003-4146-9043}}
\ShortName{Rygl et al.} 
\author[2]{Anna Bartkiewicz\orcidlink{0000-0002-6466-117X}}
\author[3]{Olga Bayandina\orcidlink{0000-0003-4116-4426}}
\author[4]{Andreas Brunthaler\orcidlink{0000-0003-4468-761X}}
\author[5]{Sandra Etoka\orcidlink{0000-0003-3483-6212}}
\author[6]{Simon Ellingsen\orcidlink{0000-0002-1363-5457}}
\author[7]{Dieter Engels\orcidlink{0000-0001-7102-6660}}
\author[8,9]{Tomoya Hirota\orcidlink{0000-0003-1659-095X}}
\author[10,4]{Arshia Maria Jacob\orcidlink{0000-0001-7838-3425}}
\author[11]{Jacco Th. van Loon\orcidlink{0000-0002-1272-3017}}
\author[12]{Alberto Sanna\orcidlink{0000-0001-7960-4912}}
\author[13]{Lucero Uscanga\orcidlink{0000-0002-2082-1370}}

\affiliation[1]{INAF- Istituto di Radioastronomia, Via P. Gobetti 101, 40129, Bologna, Italy}
\emailAdd{kazi.rygl@inaf.it}
\affiliation[2]{Institute of Astronomy, Faculty of Physics, Astronomy and Informatics, Nicolaus Copernicus University in Torun, Gagarina 11, 87-100 Torun, Poland}
\emailAdd{annan@astro.umk.pl}
\affiliation[3]{SKA Observatory, 2 Fir Street, Observatory 7925, Cape Town, South Africa}
\affiliation[4]{Max-Planck-Institut für Radioastronomie, Auf dem Hügel 69, 53121 Bonn, Germany}
\affiliation[5]{Jodrell Bank Center for Astrophysics, University of Manchester, Oxford Road, Manchester M13 9PL, UK}
\affiliation[6]{International Centre for Radio Astronomy Research, University of Western Australia, 7 Fairway, Crawley 6009, Australia}
\affiliation[7]{Hamburger Sternwarte, Universit\"at Hamburg, Gojenbergsweg 112, 21029 Hamburg, Germany}
\affiliation[8]{Mizusawa VLBI Observatory, National Astronomical Observatory of Japan, 2-12 Hoshigaoka-cho, Mizusawa, Oshu-shi, Iwate 023-0861, Japan}
\affiliation[9]{SOKENDAI (The Graduate University for Advanced Studies), 2-21-1 Osawa, Mitaka-shi, Tokyo, 181-8588, Japan}
\affiliation[10]{I. Physikalisches Institut, Universit\"{a}t zu K\"{o}ln, Z\"{u}lpicher Str. 77, 50937 K\"{o}ln, Germany}
\affiliation[11]{Lennard-Jones Laboratories, Keele University, ST5 5BG, UK}
\affiliation[12]{INAF, Osservatorio Astronomico di Cagliari, via della Scienza 5, 09047, Selargius, Italy }
\affiliation[13]{Departamento de Astronom\'ia, Universidad de Guanajuato, A.P. 144, 36000 Guanajuato, Gto., Mexico}

\abstract{Maser emission provides an unique window into astronomical sources across vast spatial scales - from tens to hundreds of astronomical units around protostars and evolved stars up to kiloparsecs in distant galaxies. 
These natural microwave amplifiers penetrate dust shells in star-forming regions, revealing the dynamics of accretion disks and outflows, trace envelopes and winds of evolved stars, map Galactic structure, while also allowing us to follow the evolution of all these systems. Owing to their compactness and brightness masers provide precise astrometry as cosmic rulers: measuring positions, structures and kinematics in dense regions, not easily accessible at other wavelengths. SKA-Mid will observe hydroxyl, methanol, and formaldehyde masers using bands 2, 5a and 5b, and later methylidyne radical masers in band 4. SKA-Mid's sensitivity and broad frequency coverage will support discoveries of new maser types and allow for simultaneous multi-transition and multi-species maser observations. In addition, bright masers can serve  in the science verification of the SKA-Mid array during the deployment phase. 
}

\begin{document}
\maketitle
\newcommand{\actaa}{Acta Astron.} 
\newcommand{\araa}{Annu. Rev. Astron. Astrophys.} 
\newcommand{\aar}{Astron. Astrophys. Rev.} 
\newcommand{\ab}{Astrobiol.} 
\newcommand{\aj}{Astron. J.} 
\newcommand{\apj}{Astrophys. J.} 
\newcommand{\apjl}{Astrophys. J. Lett.} 
\newcommand{\aplett}{Astrophys. Lett.}
\newcommand{\apjs}{Astrophys. J. Suppl. Ser.} 
\newcommand{\ao}{Appl. Opt.} 
\newcommand{\apss}{Astrophys. Space Sci.} 
\newcommand{\aap}{Astron. Astrophys.} 
\newcommand{\aapr}{Astron. Astrophys. Rev.} 
\newcommand{\aaps}{Astron. Astrophys. Suppl.} 
\newcommand{\baas}{Bull. Am. Astron. Soc.} 
\newcommand{\caa}{Chinese Astron. Astrophys.} 
\newcommand{\cjaa}{Chinese J. Astron. Astrophys.} 
\newcommand{\cqg}{Class. Quantum Gravity} 
\newcommand{\gal}{Galaxies} 
\newcommand{\gca}{Geochim. Cosmochim. Acta} 
\newcommand{\icarus}{Icarus} 
\newcommand{\jcap}{J. Cosmol. Astropart. Phys.} 
\newcommand{\jgr}{J. Geophys. Res.} 
\newcommand{\jgrp}{J. Geophys. Res.: Planets} 
\newcommand{\jqsrt}{J. Quant. Spectrosc. Radiat. Transf.} 
\newcommand{\memsai}{Mem. Soc. Astron. Italiana} 
\newcommand{\mnras}{Mon. Not. R. Astron. Soc.} 
\newcommand{\nat}{Nature} 
\newcommand{\nastro}{Nat. Astron.} 
\newcommand{\ncomms}{Nat. Commun.} 
\newcommand{\nphys}{Nat. Phys.} 
\newcommand{\na}{New Astron.} 
\newcommand{\nar}{New Astron. Rev.} 
\newcommand{\physrep}{Phys. Rep.} 
\newcommand{\pra}{Phys. Rev. A} 
\newcommand{\prb}{Phys. Rev. B} 
\newcommand{\prc}{Phys. Rev. C} 
\newcommand{\prd}{Phys. Rev. D} 
\newcommand{\pre}{Phys. Rev. E} 
\newcommand{\prl}{Phys. Rev. Lett.} 
\newcommand{\psj}{Planet. Sci. J.} 
\newcommand{\planss}{Planet. Space Sci.} 
\newcommand{\pnas}{Proc. Natl Acad. Sci. USA} 
\newcommand{\procspie}{Proc. SPIE} 
\newcommand{\pasa}{Publ. Astron. Soc. Aust.} 
\newcommand{\pasj}{Publ. Astron. Soc. Jpn} 
\newcommand{\pasp}{Publ. Astron. Soc. Pac.} 
\newcommand{\rmxaa}{Rev. Mexicana Astron. Astrofis.} 
\newcommand{\sci}{Science} 
\newcommand{\sciadv}{Sci. Adv.} 
\newcommand{\solphys}{Sol. Phys.} 
\newcommand{\sovast}{Soviet Ast.} 
\newcommand{\ssr}{Space Sci. Rev.} 
\newcommand{\uni}{Universe} 

\section{Introduction}

Maser emission is a multi-purpose tool of astrophysical research, offering unique opportunities to study dense and distant regions usually hidden at other wavelength ranges due to dust extinction. This centimetre and millimetre maser emission is relatively easily detected by single medium-sized radio telescopes and imaged by diverse radio interferometers. In favourable conditions, it appears naturally, pumped via collisions or radiation, and it is a good tracer of the specific density and temperature of the gas and dust that favour its excitation. The (often) exponential nature of maser amplification produces very bright emission, which, depending on the source geometry can be beamed both spatially and spectrally (given sufficient line-of-sight velocity coherence) into spots which can be smaller than the size of the emitting region in a velocity interval \citep[see ][]{Elitzur1992,Gray2012}. 
It is the compactness and brightness of maser spots, that enables precise astrometry to measure structures, positions, and kinematics of a vast range of astronomical sources and spatial scales - from a few astronomical units around forming and evolved stars to kiloparsecs when revealing the spiral structure of our Galaxy - making them truly cosmic rulers.

Here, we summarise the opportunities offered by the SKA-Mid array. The frequency range of this array includes two of the most abundant maser species: ground- and excited-state hydroxyl (OH) and methanol (CH$_3$OH) masers. Moreover, formaldehyde (H$_2$CO) masers can also be explored and at a later stage, the methylidyne radical (CH) will be also available. The SKA’s broad frequency range and sensitivity will be crucial for discovering and studying new 
maser transitions. They will also be important to detect and analyse weak or rare masers \citep[see, e.g., ][]{Gray2012} to reveal their origin.  Multi-band observations and the use of multiple spectral windows within one band will be of special use, when putting constraints on physical conditions required for the amplification of different transitions and/or species. In Table~\ref{tab:masers_SKAMID} we list the most common or representative maser transitions which can be studied 
with the SKA-Mid covering frequencies from 350~MHz to 15.4~GHz (see \citealt{Gray2012} for a more complete overview of maser transitions). The AA4 array will offer synthesised beam sizes, $\theta$, of 0.25"-0.71", 0.44"-0.125" and 0.024"-0.069", for bands 2, 5a and 5b, respectively, with the range depending on the chosen ($u,v$)-data weighting scheme, translating to a spatial resolution of tens to hundreds au at distances from 1 to 10~kpc. Positions of maser features (maser spots per velocity channel), can be determined to a much higher precision of about $0.5\times \theta/\mathrm{SNR}$, where SNR the signal-to-noise ratio \citep{Reid1988}. For SNRs of 50 to 100 this implies (sub-)milliarcsecond (mas) position precisions. In the following Sections, we discuss various maser science cases for SKA-Mid related to star formation and evolved stars. 

\begin{table}[]
    \centering
    \caption{A selection of maser transitions in the SKA-Mid frequency range from 350~MHz to 15.4~GHz. The more wide-spread, with $>$1000 detections, maser transitions in bold face. The columns are as follows: molecule (for OH the excited (ex) and ground (g) states are given, for \meth\ the class: I, II or unknown ``?''), transition, frequency, appearance in star-forming regions (SF) and/or evolved stars (ES), and reference of the first astronomical detection.}
    \label{tab:masers_SKAMID}
    \begin{tabular}{l l r l l l l}
    \hline
    Molecule & Transition &  Frequency &Band &   SF/ES & Reference \\
             &            &        (GHz)& &     & \\
    \hline
    \meth\ cI?& $J_k$= 1$_1$--1$_1$ A$^\pm$($v_t$=0)&0.83427 & 1&  y/& \cite{ball1970}\\
    {\bf g-OH} & $^2\Pi_\mathrm{3/2}$ $J$= 3/2, $F$=1--2 &   1.61223 &2&  y/y & \cite{wilson1968}\\
    {\bf g-OH} & $^2\Pi_\mathrm{3/2}$, $J$= 3/2, $F$=1--1 &   1.66540 &2&   y/y & \cite{weaver1965}\\
    {\bf g-OH} & $^2\Pi_\mathrm{3/2}$, $J$= 3/2, $F$=2--2 &   1.66735 &2&   y/y & \cite{weaver1965}\\
    {\bf g-OH} & $^2\Pi_\mathrm{3/2}$, $J$= 3/2, $F$=2--1 &   1.72053 &2&  y/y & \cite{gosswm1968}\\    
    CH & $^2\Pi_\mathrm{1/2}$ $J$= 1/2, $F$=0--1& 3.26379&4&y/&\cite{Rydbeck1973} \\
    CH &$^2\Pi_\mathrm{1/2}$ $J$= 1/2, $F$=1--1& 3.33548&4&y/&\cite{Rydbeck1973}\\
    CH &$^2\Pi_\mathrm{1/2}$ $J$= 1/2, $F$=1--0& 3.34919&4&y/&\cite{Rydbeck1973}\\
    ex-OH & $^2\Pi_\mathrm{1/2}$, $J$= 1/2, $F$=0--1  &  4.66024 &5a& y/y & \cite{Thacker1970}\\
    ex-OH & $^2\Pi_\mathrm{1/2}$, $J$= 1/2, $F$=1--1  &  4.75065 &5a& y/y & \cite{Garnder1971}\\
    ex-OH &$^2\Pi_\mathrm{1/2}$, $J$= 1/2, $F$=1--0  &  4.76556 &5a&   y/y & \cite{Zuckerman1968}\\  
    \form\ & $J_{K_a,K_c}$= 1$_{1,0}$--1$_{1,1}$ &  4.82966 &5a&  y/ & \cite{Downes1974}\\
    \meth\ cI?& $J_k$= 3$_1$--3$_1$ A$^\pm$($v_t$=0)&  5.00532 &5a &  y/& \cite{robinson1974}\\
    ex-OH &  $^2\Pi_\mathrm{3/2}$, $J$= 5/2, F=2--2  &  6.03075 &5a&   y/y & \cite{Yen1969}\\
    ex-OH &  $^2\Pi_\mathrm{3/2}$, $J$= 5/2, F=3--3  &  6.03509 &5a&  y/y & \cite{Yen1969}\\
    \meth\ cII& $J_k$= 17$_{-2}$--18$_{-3}$ E($v_t$=1) & 6.18113 &5a &   y/ &\cite{Breen19}\\ 
    {\bf CH$\mathbf{_3}$OH cII}& $J_k$= 5$_1$ -- 6$_0$ A$^+$($v_t$=0) &  6.66851 &5a &  y/n &\cite{Menten1991}\\
     \meth\ cII& $J_k$= 12$_4$--13$_3$ A$^-$($v_t$=0) & 7.68223 & 5a& y/ &\cite{Breen19}\\ 
    \meth\ cII& $J_k$= 12$_4$--13$_3$ A$^+$($v_t$=0) & 7.83086&5a&   y/ & \cite{Breen19}\\
     HC$_3$N & $J$= 1--0 & 9.09811 &5b& y/ & \cite{Morris1976}\\
    \meth\ cI& $J_k$= 9$_{-1}$ -- 8${_2}$ E2($v_t$=0)&  9.93620&5b &   y/ & \cite{slysh1993}\\
    CH${_3}$OH cII& $J_k$= 2$_1$ -- 3$_0$ E($v_t$=0) &  12.17859& 5b& y/n &\cite{Batrla1987}\\ 
    ex-OH & $^2\Pi_\mathrm{3/2}$, $J$= 7/2, $F$=4--4 &13.44137 & 5b & y/n & \cite{Turner1970}\\
    \form\ & $J_{K_a,K_c}$= 2$_{1,1}$--2$_{1,2}$ & 14.48848 &5b&   y/n &\cite{Wilson1982}\\ 
    \hline
    \end{tabular}
\end{table}

\section{Star formation}

The evolution and formation of stars is directly connected not only to disk and planet formation, but also to galaxy evolution, as stars shape the Galactic interstellar medium (ISM) through chemical enrichment, UV-radiation, and pressure and turbulence injections through molecular outflows, stellar winds, and supernovae. To date, star formation, and in particular, high-mass star formation, is not yet fully understood, as observational data of the immediate surroundings of high-mass young stellar objects (HMYSOs) are extremely difficult to obtain. Their formation phase  is very short \citep[$\sim10^5$ years;][]{Motte2018} and occurs while they are still deeply embedded in the natal molecular cloud. 
Additionally, there are relatively few HMYSOs, and, with few exceptions, they lie at distances of several kpc or more, concentrated in the Galactic plane. To study their physical properties and kinematics at the relevant spatial scales of thousands to a few au, high angular resolution is required.

The onset of high-mass star formation occurs in preferential locations such as filaments and filament hubs, where various star-forming cores may compete for accreting gas and dust \citep[see][and references therein]{Motte2018}. After the formation of a gravitationally bound core, collimated jets and wide-angle bipolar outflows are observed together with circumstellar disks (e.g. \citealt{kuiper2010,anglada2018}). During their evolution, HMYSOs are thought to acquire the bulk of their mass through episodic bursts of accretion \citep{Meyer2021}.
Only infrared and radio wavelength ranges allow one to explore these distant and very dense regions of HMYSOs. In this respect, cosmic masers arising in the vicinity of newly born stars are unique tools to study the structure and kinematics of the surrounding gas.

In particular, class II methanol masers are considered an unique tracer of high-mass star formation \citep{Menten1991}, and thought to arise through radiative pumping in the disk and envelope surrounding the HMYSO. Also, ex-OH maser emission is originating in similar regions and conditions \citep{kobak2025} via radiative pumping. Only a few HMYSOs have formaldehyde masers; these masers are on average weaker than the aforementioned species and are thought to arise in disks around HMYSOs \citep{araya2007_review}. Class I methanol masers are excited in non-dissociative post-shocked gas \citep{Leurini2016}, and typically observed toward large-scale outflows, far away from the HMYSO, where they impact with the ambient gas of the parental cloud \citep{Cyganowski09}. In summary, given that different maser species arise in different regions near the forming star, multi-maser observations are necessary to obtain a comprehensive picture.

Masers can also be used to measure the magnetic field strength and orientation through Zeeman splitting. This is strongest for the OH radical but can be measured for most masers at
high enough angular and velocity resolution (see, e.g., \citealp{fish:2003} for OH and \citealt{Dallolio2020} for methanol). Observations of the magnetic field at au-scales in HMYSOs provide important constraints on star formation theories. Zeeman splitting and magnetic fields, including those from masers, are discussed in \cite{Bourke01.2026.SKA,Robishaw01.2026.SKA}.

\subsection{Maser surveys and serendipitous detections}
\begin{table}[]
\begin{threeparttable}
\centering
\caption{Untargeted (blind) 6.7~GHz methanol maser surveys and their detections. The SKA-Mid estimate is given for 10\,min integration time towards NGC\,6334I, using a bandwidth for sensitivity of 0.08\,$\mathrm{km\,s^{-1}}$ and a Briggs robust=$-1$ weighting. 
}
    \label{tab:masers_surveys}
    \begin{tabular}{l c c c l} 
    \hline
Survey & Sensivity (1$\sigma$) & Galactic Longitude & \# detections/new &References\\
 & (Jy) & & & \\
\hline
26-m Mt Pleasant & 0.9  & 325$^{\rm o}-335^{\rm o}$ & 50/26 & \cite{ellingsen1996} \\
ATCA & ~~~~0.16$^a$ & 330.8$^{\rm o}-339.8^{\rm o}$ & 57/21 & \cite{caswell1996}\\
25-m Onsala & 0.6 & 35$^{\rm o}-220^{\rm o}$ & 11/4 & \cite{pestalozzi2005} \\
32-m Torun & 0.6 & 20$^{\rm o}-40^{\rm o}$ & 100/26 & \cite{szymczak2002} \\
Arecibo & 0.27  & 35.2$^{\rm o}-53.7^{\rm o}$ & 86/48 & \cite{pandian2007a} \\
64-m Parkes &  0.17 & 345$^{\rm o}-0^{\rm o}-6^{\rm o}$ & 183/48& \cite{caswell2010} \\
64-m Parkes & 0.17 & 6$^{\rm o}-20^{\rm o}$ & 119/42 & \cite{green2010}\\
64-m Parkes & 0.17 & 330$^{\rm o}-345^{\rm o}$ & 198/80 & \cite{caswell2011}\\
64-m Parkes & 0.17 & 186$^{\rm o}-330^{\rm o}$ & 207/89 & \cite{green2012}\\
64-m Parkes & 0.17 & 20$^{\rm o}-60^{\rm o}$ & 265/64 & \cite{breen2015}\\
VLA (D-conf.) & 0.018$^a$ & $-2^{\rm o}-60^{\rm o}$ & 554/84 & \cite{nguyen2022}  \\
SKA-Mid AA4 & 0.0051$^a$ & & & \\
\hline
\end{tabular}
\begin{tablenotes}
\item[$^a$] unit: Jy/beam (interferometric data)
\end{tablenotes}
\end{threeparttable}
\end{table}

Systematic surveys of the Galactic plane are the main channel to increase the number of maser detections in our Galaxy. To date, the most complete maser database can be found on \url{https://maserdb.net/}. 

\subsubsection{Methanol maser surveys - towards a complete Galactic HMYSOs sample}

The 6.7\,GHz methanol maser transition has often been used in blind surveys due to its brightness and its property of tracing high-mass star formation. Table~\ref{tab:masers_surveys} lists the untargeted (blind) surveys of the 6.7~GHz methanol maser line. 
It is clear that more sensitive surveys have systematically discovered new methanol maser targets, and more remain to be discovered. For example, the commissioning of the multi-beam C-band receiver, at the Parkes 64-m radio telescope in Australia, increased by roughly 30\% the total number (972) of Galactic methanol maser sources known across a large Galactic longitude range of $186^{\rm o}$ through 0$^{\rm o}$ to 60$^{\rm o}$  
(Methanol Multi-Beam survey, MMB; \citealp{green2009},
\citealp{breen2015}). Although the Galactic longitude range between 20$^{\rm o}$ and $60^{\rm o}$ had been extensively searched for new methanol maser sources before with different telescopes, the MMB survey still reported 64 new detections in addition to the 265 known sources in that region. 

Today, we know that weaker masers also exist: the Arecibo radio telescope targeted 107 massive cores found in the {\em Herschel} infrared Galactic Plane Survey \citep[Hi-GAL][]{Molinari:2010} and reported 22 new methanol maser sites out of 37 detections, with a median peak flux density of only  0.07~Jy  \citep{olmi2014}. Still, the question outlined by \cite{breen2013} remains valid: {\it While it is widely accepted that low-mass YSOs are unable to produce class II methanol masers, what is the lower mass limit on the stars that can produce them?}

The large collecting area of SKA-Mid will be crucial to reliably discover weak, sub-mJy, masers. For comparison, we list in Table~\ref{tab:masers_surveys} also the SKA-Mid sensitivity at the 6.7~GHz methanol frequency assuming Briggs robust=$-$1 ($u,v$)-data weighting\footnote{With a more natural weighting, the sensitivity improves almost by a factor 2 at the cost of a similar reduction in angular resolution.}. The SKA-Mid's increase in sensitivity will directly impact the number of future methanol maser detections and enhance the completeness of various maser species catalogues due to the wide field of view and instantaneously covered bandwidth to include various maser transitions, as discussed in the next Section. 

\subsubsection{Surveys of various maser species and transitions}
Systematic surveys of various maser transitions from Table~\ref{tab:masers_SKAMID}, including OH, 6.7 and 12.2~GHz CH$_3$OH, as well as CH and H$_2$CO masers, will be of great importance. In 10 minutes using  robust=$-$1 Briggs weighting, SKA-Mid AA4\footnote{The SKA-Mid sensitivity calculator, \url{https://sensitivity-calculator.skao.int/\#/mid-calculator}, exists for bands 2, 5a, and 5b, but not yet for band 4 (CH).} can reach toward NGC\,6334I $\sim$9~mJy/beam for g-OH masers and $\sim$5 ~mJy/beam for ex-OH, \meth\ and  H$_2$CO masers. Using a frequency setup with multiple zoom windows per band, SKA-Mid could simultaneously detect, for example, H$_2$CO  and ex-OH masers together with 6.7\,GHz \meth. Planned SKA-Mid Galactic plane surveys, of band 5b continuum \citep{Traficante01.2026.SKA} and of the ionised medium \citep{Karska01.2026.SKA}, will have the potential to detect various maser transitions in a systematic and blind fashion, albeit with a coarser velocity resolution and a limited sensitivity per velocity channel, yielding important population statistics of the covered transitions. Also new maser flares may be caught by these surveys, as simultaneously with the episodic 6.7\,GHz methanol maser flares (discussed in Section~\ref{sec:sf-time}) a number of new maser transitions were reported to brighten connected to those flaring events (e.g. \citealp{Breen19}). 

In brief, with more sensitive and complete maser catalogues containing more maser transitions, we can expect the following: 
\begin{itemize}
    \item to improve on the evolutionary scenario of maser excitation in our Galaxy (e.g., \citealp{billington2020});
    \item to investigate the exclusive association of specific maser transitions and HMYSOs, such as for the 6.7~GHz class II methanol line (e.g., \citealp{breen2013});
    \item to reveal less known or rare maser transitions, to locate them, and to study their environments \citep{Breen19};
    \item to compare maser populations across different galactic environments and structures \citep{caswell2011}.
\end{itemize}

Notably, the combination of improved maser models with the detection of different maser transitions overlapping spatially (co-propagating in space) will possibly allow us to constrain physical conditions and molecular abundances within star-forming regions. \cite{kobak2025} reported coincidence of methanol and ex-OH maser cloudlets within 200 au separation of cloud centres and peak velocities within 0.7\,\kms. For a typical distance of 3\,kpc this would require $\sim$ 100 mas to resolve such a separation by 5 synthesised beams, which can be easily achieved with SKA-Mid. These objectives will further require a sensitivity (3$\sigma$) better than 10~mJy~beam$^{-1}$ per 0.1~km~s$^{-1}$ spectral channel, and flexible frequency settings to cover multiple frequency ranges simultaneously per band. 

Together with blind Galactic plane (maser) surveys, targeted surveys of homogeneous samples of young stars, based on evolutionary stage or particular characteristics, such as the presence of a jet or outflow, will be seminal to statistically assess and correlate those characteristics with maser properties. The latter can then, through models, be used to understand the physical processes at work on circumstellar distances of less than a thousand au. 

\subsubsection{Combining maser- and continuum-emission surveys}
Aside of comparing different maser transitions and species, maser emission in combination with continuum emission is also important to consider. The interferometric observations by \cite{Hu16} showed that at least a third of all 6.7~GHz methanol masers have a radio continuum counterpart, which most likely arises from free-free emission from a young H~{\small II} region. The probability of finding such a radio continuum counterpart increased with the luminosity of the maser. Observations with high-angular resolution using very-long baseline interferometers (VLBI) of the 6.7\,GHz transition have shown that compact maser cloudlets of a few au in size are made of spots grouped in ordered velocity gradients, where gas slowly moves by few km\,s$^{-1}$ only \citep[e.g.,][]{sanna2010a,moscadelli2011mas,bartkiewicz2024}. These coherent motions of quiescent gas can be used to trace circumstellar regions and describe physical processes happening near a young star similarly as it has been done for water masers in, e.g., the recent Protostellar Outflows at the EarliesT Stages (POETS) project. The survey has imaged the inner portion of the wind (on scales of 10$-$100~au) in a significant sample of 37 luminous YSOs using the VLA and VLBA \citep[e.g.,][]{moscadelli2020,sanna2018}. The results indicate that weak radio continuum emission ($\leq$\,100\,$\mu$Jy), associated with ionised winds and jets, should always be detected near water maser emission from young stars; in these sources, magneto-hydrodynamic disk winds appear a major driver for the dynamics of the innermost water masers.

SKA-Mid sensitive and simultaneous observations of the radio continuum (at nano-Jy) and methanol masers (at sub-mJy), will allow us to image the protostellar environments much more completely (e.g. not only the brightest peaks as in VLBI), deeply and with a higher detail than the large-scale interferometric methanol surveys, potentially explaining the precise origin of 6.7~GHz methanol masers excitation. Furthermore, with the extension of SKA-Mid to 25\,GHz, water masers will also be covered and we expect to double the number of water maser detections with respect to a typical VLBI threshold of some 10\,mJy\,beam$^{-1}$ (\href{https://www.skao.int/sites/default/files/documents/d38-ScienceCase_band6_Feb2020.pdf}{Memo 20-01: SKA1 Beyond 15GHz}).

\subsubsection{Opportunities to survey unsurveyed masers}

H$_2$CO masers are much rarer and less studied than OH and \meth\ masers. \cite{pratap1994,araya2008,chen2017brightest} (and references therein) detected H$_2$CO masers in star-forming regions using the Very Large Array (VLA) or large-aperture single dishes. These masers appear to trace a very early phase of the star formation process, but the reason for their rarity is not clear. The conditions required to excite the H$_2$CO molecule may be specific and short-lived. With SKA-Mid, deeper H$_2$CO maser surveys could be performed simultaneously with other band 5a and 5b masers (see Table~\ref{tab:masers_SKAMID}) and continuum emission, with the potential to constrain physical conditions (density, temperature) of H$_2$CO maser excitation in the HMYSO environments. The same can be done for cyanoacetalyne (HC$_3$N) masers, which are even rarer -  with only one single detection towards the Sgr\,B2 star forming region \citep{Morris1976,Brown1985}.

As the simplest carbyne, CH plays a central role in interstellar chemistry, but is best known for its strong correlation with molecular hydrogen (H$_2$). Optical absorption studies have shown that CH column densities closely track those of H$_2$ and visual extinction in diffuse and translucent clouds \citep{federman1982measurements, sheffer2008, Weselak2019}, making it a widely used proxy for tracing H$_2$, which is otherwise difficult to measure directly \citep{Karska01.2026.SKA}. CH shows weak maser emission at 3.264, 3.335, and 3.349~GHz \citep[see Table~\ref{tab:masers_SKAMID}, e.g.][]{Zuckerman1975, Rydbeck1976}. Early attempts to explain this maser behaviour proposed a simple pumping cycle involving collisional excitation (primarily by H and H$_2$) to higher rotational states, followed by radiative decay back to the ground state \citep{Bertojo1976, Elitzur1977}. Recent findings firmly establish that CH ground-state transitions represent a weak but widespread astrophysical maser phenomenon; their excitation involves a delicate balance of collisional and radiative processes whose relative importance depends on the local environment (e.g. \citealp{Bujarrabal1984,Jacob2021,Jacob2023}). Even though the CH masers are weak, they offer a uniquely sensitive diagnostic of the physical and chemical conditions during the earliest phases of molecular gas evolution that precede star formation. They emerge as an important band 4 probe -- one that should be actively considered for scientific exploration with SKA-Mid.

\subsection{Gas kinematics in the close vicinity of the young star}

\begin{figure}[!h]
    \centering
    \includegraphics[width=0.95\linewidth]{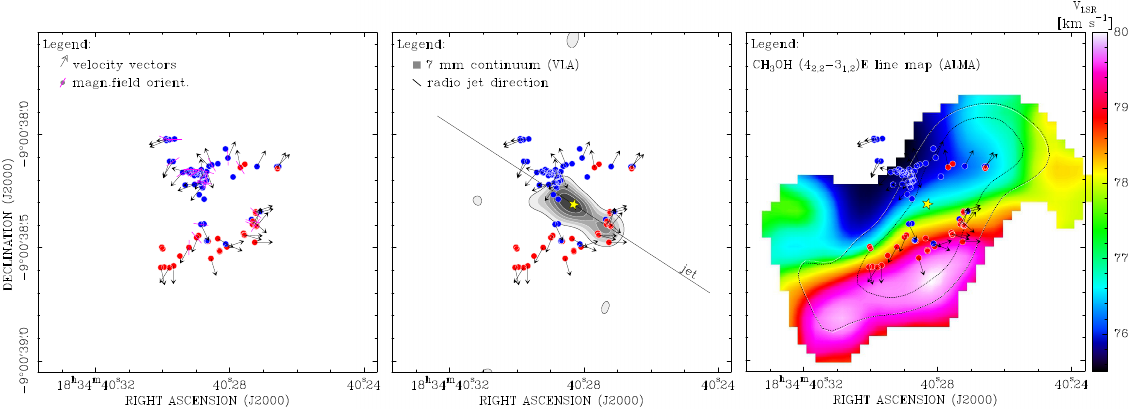}      
    \caption{6.7~GHz methanol masers towards HMYSO G023.01$-$00.41. {\it Left panel:} distribution of the methanol maser cloudlets (dots) mapped with the EVN (blue and red colours indicate blue- and red-shifted velocities). The velocity vectors are suggestive of expansion from a common centre. {\it Middle panel:} similar to the left panel, but  with the YSO position (yellow star) and the 7\,mm continuum tracing an ionised thermal jet. {\it Right panel:} similar to the left panels, but overlaid with a first moment map from (thermal) methanol gas observed with ALMA. This comparison shows that the maser emission arises from an extended plateau of methanol gas tracing a large molecular disk, from where the compact maser cloudlets are blown away. Figure adapted from \cite{sanna2015, sanna2019}.}
    \label{fig:G2301}
\end{figure}

Sensitive, systematic and multi-transition VLBI and interferometric studies of selected regions can reveal the au-scale gas kinematics around HMYSOs, where gas can be channelled onto the accretion disk, or be ejected through molecular jets. For example, European VLBI Network (EVN) 6.7~GHz methanol maser line observations of HMYSO G023.010$-$00.411 
motivated further radio to infrared studies using VLA and Atacama Large Millimeter/submillimeter Array (ALMA), revealing that this object has a rotating molecular disk undergoing infall while driving a molecular jet as illustrated in Fig.~\ref{fig:G2301} \citep{sanna2010b,sanna2014,sanna2019,sanna2021}. A sensitive array is needed to explore the detailed structures of masing regions, as these can have a wide diversity of distributions and luminosities \citep[see e.g.][]{bartkiewicz2009,fujisawa2014}. As such, higher sensitivity imaging with SKA-Mid will allow for a more complete study, including weaker maser cloudlets and more extended maser emission. Imaging weaker maser features may reveal new morphologies which will help to understand the underlying three-dimensional structure of pumped gas (see, e.g., \citealp{Bartkiewicz2005} and references therein). This requires good imaging quality, and consequently a good ($u,v$)-coverage, as offered by SKA-Mid. The combination of VLBI and intermediate angular resolutions is essential as up to 50-70\% flux density is filtered out for methanol maser emission when observed with VLBI \citep{bartkiewicz2016}. Illustrative comparisons of spatial filtering are shown in a number of methanol masing sources imaged with the VLA \citep{Cyganowski09}, the Multi-Element Radio Linked Interferometer Network (MERLIN) \citep[][]{pandian2011} and the EVN \citep[][]{bartkiewicz2016}. 

So far, the limited sensitivity and spatial scale coverage of VLBI facilities suggested that masers can only
provide a partial view of the gas they arise from. This is because VLBI maser maps typically appear patchy and only a limited number of (relatively) bright spots can be detected: these brightest spots mark parcels of gas where amplified emission is beamed almost along the line-of-sight. As a consequence, maser studies of individual objects have not been conclusive by themselves, and a consistent interpretation of the maser environment has required the use of thermal line emission maps in all cases (e.g. \citealt{sanna2010b, moscadelli2019alma, sanna2021}). We note here that the absence of observed maser emission does not exclude the presence of masers, as beaming conditions may not be favourable for our line of sight. Evidence for fainter and more diffuse maser emission was found for methanol and OH masers (e.g., \citealt{harvey-smith2006}) giving a more complete picture of the gas,  than that provided only by the brightest, more compact maser spots. The behaviour of maser intensity with spot size contains information about whether the maser is excited in a quiescent or shocked region \citep{Richards2011}. 
With the exponential increase in spectroscopic sensitivity, deep SKA-Mid observations will be transformational for studies of faint maser emission, both faint maser peaks and faint emission in the line wings, in HMYSOs.

In particular, maser studies maximise their information when the proper motions of individual maser cloudlets are also measured, thus providing the three-dimensional motion of gas at the 0.1~km~s$^{-1}$ accuracy. The majority of 6.7~GHz masers should be excited in the outer disk layers (or inner outflow cavities) at a few 1000~au that are slowly swept away from the star, while this maser species is quenched in the innermost 100s of au at gas densities exceeding $\rm 10^8~cm^{-3}$.  Multi-epoch data will permit to observe the evolution in time, and trace those expanding motions (e.g. \citealt{sanna2010b,bartkiewicz2020}) to study the physical properties of the slow disk winds traced by methanol masers. This would require at least two mas-scale accuracy observations of the same methanol maser target over the baseline of a few years, taking into account typical distances and average speeds of the methanol gas they are associated with. As highlighted earlier, this is easily achievable with SKA-Mid for high SNR maser features. By directly comparing these measurements with simulations \citep{koelligan2018,oliva2023}, proto-stellar growth may be better constrained. Eventually, these data sets can be distilled into short-featured movies explaining the kinematic evolution of methanol gas in nearby sources, both for scientific and educational means.

\subsection{Time domain maser studies - maser flares}
\label{sec:sf-time}

The 6.7~GHz methanol maser line was recently found to be sensitive to episodic, disk-mediated accretion in HMYSOs. The first manifestation of such accretion events is a sudden 6.7\,GHz methanol maser flare, an intense and short-lived flux density surge. Simulations of \cite{Meyer19} showed that massive fragments formed through gravitational fragmentation in the outer accretion disk  regularly generated sudden increases of the accretion rate when fragments migrate towards the protostar and produced violent, accretion-driven, luminosity bursts in the protostellar light curves. HMYSOs can acquire 40-60\% of their mass in this phase \citep{Meyer2021}. Since the first discovery in 2016, three accretion bursts in HMYSOs have been confirmed and in all these cases, the 6.7~GHz methanol maser line flared: S255IR \citep{Caratti17}, NGC\,6334I \citep{Hunter17}, G358.93$-$0.03 \citep{Stecklum21}, and the NIR surge and methanol maser flare were associated. For a few more accretion burst candidates the evaluation is ongoing, as maser flares can have also other origins as discussed in \cite{Gray2020}. As NIR monitoring is much more challenging than 6.7\,GHz maser monitoring, which is done by many medium-sized radio antennas worldwide, the 6.7\,GHz methanol maser became the primary accretion burst indicator. Accretion bursts are also found to be followed by delayed flares in the collisionally pumped 22~GHz water maser transition (e.g. \citealt{Bayandina22}), which can trace the new ejection episodes following the accretion episode \citep{Meyer2021,Gray2024}.

\begin{figure}
    \centering
    \includegraphics[width=0.47\linewidth]{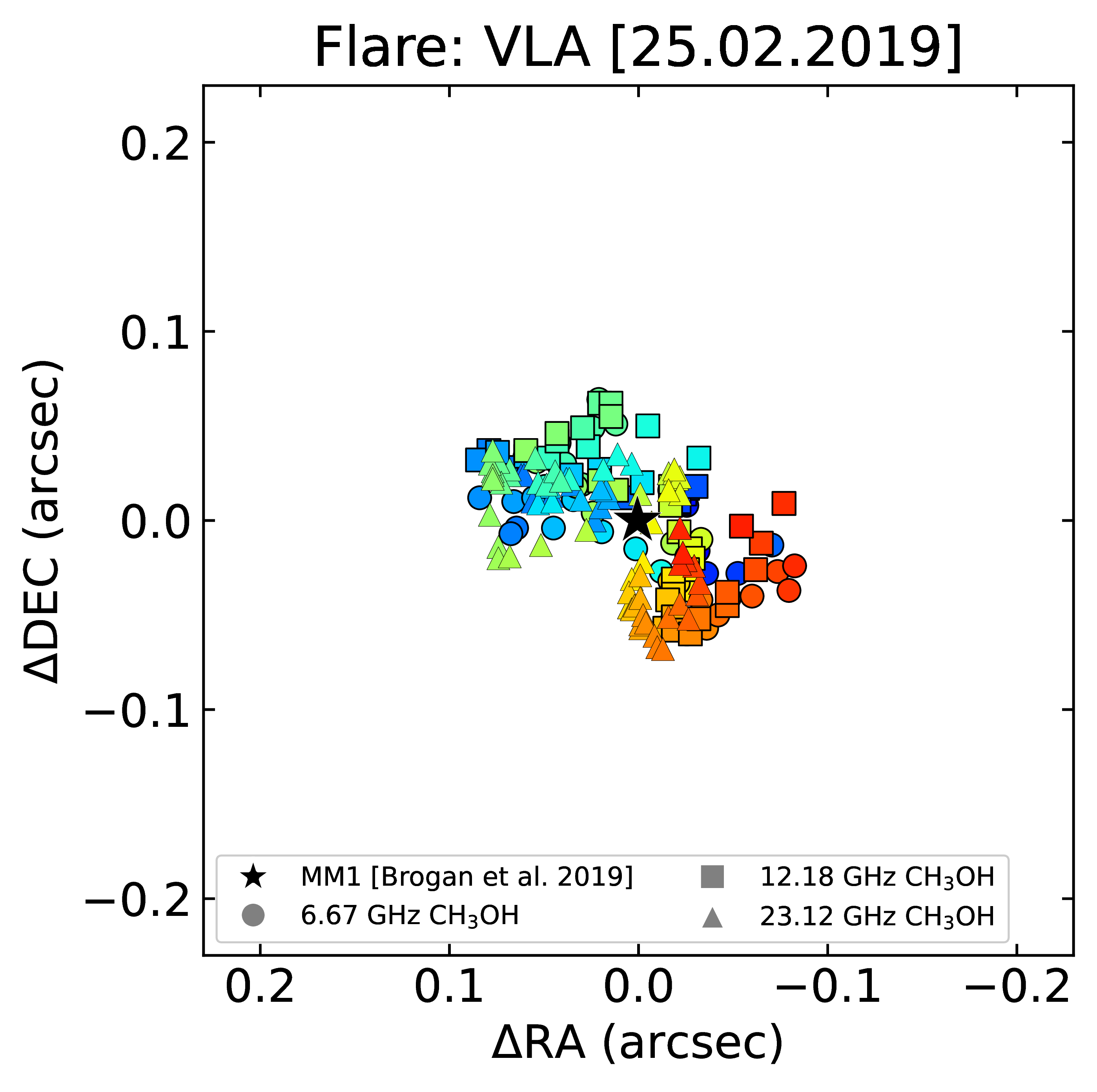}
    \includegraphics[width=0.47\linewidth]{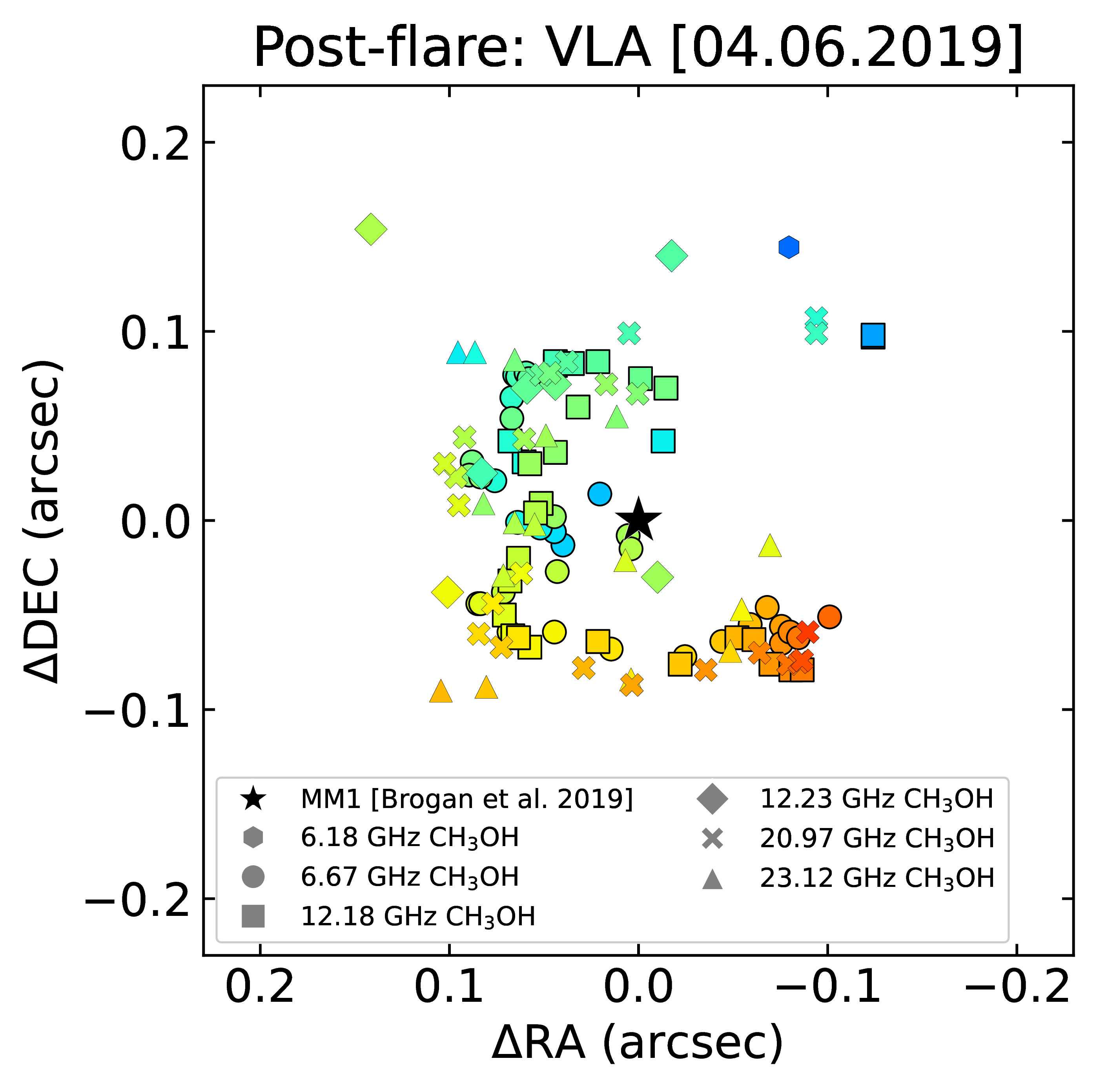}
    \caption{Comparison of spatial distribution of the methanol maser emission in G358.93-0.03 detected with the VLA by \cite{Bayandina2022} at the flare and post-flare epochs, shown in the left and right panels, respectively. Figure adapted from \cite{Bayandina2022}. } 
    \label{fig:bayandina}
\end{figure}

To improve our understanding of the accretion process in HMYSOs; finding new accretion bursts is a high priority. To utilise the potential of masers as tracers of accretion bursts, \href{https://masermonitoring.org}{the Maser Monitoring Organisation (M2O)} was established in 2017 \citep{burns2024}. An example of the 6.7~GHz methanol maser flare in HMYSO G358.93$-$0.03 and its subsequent follow-up studies shows the importance of international collaboration and use of diverse instruments. The outburst was initially detected in the Ibaraki 6.7 GHz class II methanol maser monitoring program\footnote{The iMet website is available at \url{http://vlbi.sci.ibaraki.ac.jp/iMet/}.} \citep[iMet, ][]{Sugiyama2019} and later, the coordinated follow-ups by M2O led to unique discoveries. Single-dish observations revealed that the methanol maser flare was happening not only at 6.7~GHz but also in more than 30 other transitions over a wide frequency range from 6.18 to 361.2~GHz (e.g. \citealt{Breen19, MacLeod19}). Multi-epoch VLA and VLBI observations were able to follow the propagation of the heatwave of the burst through the accretion disk as the region of the optimal physical parameters for maser excitation moved from the inner to outer radii (\citealt{Bayandina2022, Burns2023}). Figure~\ref{fig:bayandina} shows the wide band VLA observations with which the propagation of the heatwave was traced not only with the bright 6.7~GHz maser (which was the only tracer available for the VLBI observations) but also with weaker, newly discovered masers \citep{Chen20, Bayandina2022}. An innovative approach of combining all the VLBI images obtained at each stage of the heatwave propagation, so-called ``heatwave mapping'', produced the highest-ever-archived-resolution image of an accretion disk around an HMYSO and proved the presence of spiral arms in it \citep{Burns2023}. Follow-up studies of the 22~GHz water maser emission in G358.93$-$0.03 revealed that the accretion burst affected not only its immediate vicinity (i.e. accretion disk) but also the entire host region \citep{Bayandina22, Miao2024}. 

SKA-Mid will be transformative for the time domain observations of masers thanks to its fast survey speed, high sensitivity, and frequency agility, especially for confirming and characterising maser flares associated with accretion bursts in HMYSOs. For example, observerving NGC\,6334I at 6.7~GHz, the rms per spectral channel (assuming 0.15~km~s$^{-1}$ channels, 1-hour integrations, and robust=1 Briggs weighting) will be ca.~0.6~mJy with SKA-Mid, representing a factor$\sim$5 improvement in sensitivity with respect to the VLA. Similarly for continuum observations with a ~GHz bandwidth, SKA-Mid achieves a sensitivity of 0.8~$\mu$Jy compared to 4~$\mu$Jy for the VLA. In addition, SKA-Mid’s long baselines provide five times better angular resolution of $\sim$60~mas at 6.7~GHz with a natural ($u,v$)-data weighting than the VLA in A-configuration. We expect SKA-Mid to provide important constraints on how accretion bursts affect the immediate star-forming environment: 

\begin{itemize}

\item To confirm accretion bursts: precisely locate flare sources detected by single-dish monitoring programmes and provide coordinates for VLBI follow-up with sub-mas astrometric accuracy and detect subtle changes in maser structure and flux that are beyond the reach of current instruments. Absolute astrometric registration will be possible for simultaneously observed maser species, and for HMYSOs with bright and compact enough continuum emission, this could be extended to other maser species.

\item To characterise maser emission: investigate the spatial structure of masers, including weak transitions not accessible to present-day interferometers (e.g. faint masers of only a few hundred mJy) and conduct semi-simultaneous observations across all SKA-Mid frequency bands (see Table~\ref{tab:masers_SKAMID}), providing a comprehensive view of maser excitation conditions and enabling multi-transition diagnostics. The wide instantaneous bandwidth holds the potential to discover new masers in flaring sources \citep[e.g.,][]{Chen20}. 

\item To monitor burst evolution: track the progression of accretion bursts in detail, even for weak or short-lived maser features, and evaluate changes in maser emission across multiple transitions to infer the physical conditions within the accretion disk and to test theoretical predictions of heatwave propagation. 
\item To assess environmental impact: detect and monitor changes in continuum sources associated with maser flares at higher sensitivity than currently possible and characterize the natal regions of maser flares through thermal spectral lines within SKA-Mid’s frequency coverage. 
\end{itemize}

Even before the full baseline capabilities are available, SKA-Mid in its AA* configuration will already provide critical new insights, because the most important requirements for maser flare follow-up are: rapid response, astrometry of the flare sources, multi-frequency observations, and monitoring. There is also a high potential of commensality as the high brightness of maser flares makes them easily detectable in other, not maser-targeted observations, even in fast-scanning sky surveys. 
Extending SKA-Mid’s frequency coverage beyond 15.4~GHz will allow to include the 22~GHz water maser to trace not only the accretion disk but also the episodic ejections and jet activity triggered by bursts.

\subsection{Star-formation studies through statistics and multi-frequency observations}
Having many detections of several maser species, each with its own characteristic pumping conditions, one may try to perform a statistical analysis to constrain physical properties of the maser-emitting regions surrounding the YSO, as was done by \cite{Breen2018} for methanol, ex-OH and water masers. The presence of a particular maser transition associated with an astrophysical object requires three conditions to be met: 
\begin{itemize}
\item Sufficient gas-phase abundance of the relevant molecule.
\item Physical conditions in the molecular gas that produce inversion for the relevant transition.
\item Sufficient line-of-sight velocity coherence along the direction between the astrophysical object and the Earth.
\end{itemize}
This means that any transition which is widely observed to show strong maser emission must be from a common molecule for which the physical conditions of molecular gas readily produce a population inversion and does not require a highly specific geometry of the source with respect to the observer. Conversely, transitions which are rarely observed to show maser emission must fail one or more of the above requirements.

\begin{figure}
    \centering
    \includegraphics[width=0.82\linewidth]{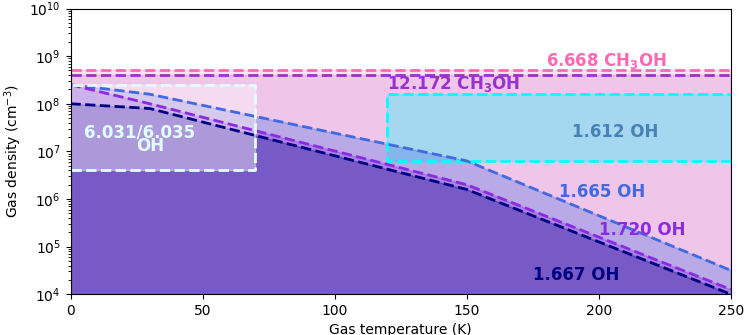}
    \caption{Estimates of the gas density and temperature parameter space for a number of OH and \meth\ maser transitions in the SKA-mid frequency range. Figure after \cite{green2007}.}
    \label{fig:cragg}
\end{figure}

Ideally, sensitive high-resolution observations of a range of molecular transitions can reveal information on the physical conditions in the molecular gas at the observed scales. For example, Fig.~\ref{fig:cragg} summarises the gas temperature and density for \meth\ \citep{cragg2005} and OH  \citep{cragg2002} maser excitation based on models. In practice, this is a very challenging task, both observationally and theoretically. Modern, sensitive interferometers that have large instantaneous frequency coverage and high spectral resolution, such as ALMA and in the future SKA-Mid, provide a wealth of observational information, but inference of physical conditions from that data will require significant effort to improve our theoretical models of molecular emission. In particular, maser models need to cover a realistic range of excitation conditions, and include competition or co-propagation between masing (and other) transitions and source geometry \citep[see][and references therein]{Gray2012}. 

The two molecules which show the greatest potential for multi-transition studies with the Mid-SKA are, above all, OH and CH$_3$OH. Each of these has multiple maser and thermal transitions in the SKA frequency range and has been detected towards more than 1000 Galactic sources. Moreover, the methylidyne radical (CH) is one of the most ubiquitous molecules in the interstellar medium (ISM), observed across a wide range of environments \citep[e.g.,][]{Rydbeck1976}. 

Large maser catalogues are also important to improve the pumping models for a single maser species. For example, for 6.7\,GHz methanol, we know that masers arise in both extended and compact regions \citep{bartkiewicz2016}, but with the broad spatial (and spectral) coverage of SKA-Mid, such studies can be extended to many more star-forming regions (and other maser transitions). By associating kinematic (or parallax) distances to these regions, the physical scales of masing clouds can be determined in a statistically meaningful manner. At the same time, the maser luminosity function can be derived to much lower luminosities than in previous works \citep{pandian2009,green2011_H_abs} given SKA-Mid's sensitivity. Maser luminosity functions can then be used to understand and predict the presence of a particular maser species in nearby galaxies or Milky Way satellites, and also the total number of masing sources expected in the Milky Way itself.

\subsection{Astrometry of HMYSOs on the far side of the Milky Way}
In the last decades, masers in star-forming regions have been an invaluable tool to map the spiral structure of the Milky Way, as radio observations provide an unobscured view into the dense, high-extinction regions in spiral arms. 
VLBI arrays in the northern hemisphere have measured very accurate trigonometric parallaxes to about 200 maser sources in HMYSOs with accuracies similar or even exceeding the astrometric accuracy of Gaia (e.g. \citealt{rygl2012, reid2019, Hirota2020}). These helped to locate the positions and properties of the spiral arms in the Milky Way. While there have been some parallax measurements from the southern hemisphere in recent years \citep{Hyland2023}, surveys of a large number of sources with current telescopes will be a challenge. 

SKA-Mid, with baselines of up to 150~km alone will not be able to provide the necessary accuracy to measure parallaxes of masers on a Galactic scale. However, an alternative method exists to measure accurate distances to sources which are very far away: the three-dimensional (3D) kinematic distance \citep{Reid2022}. Here, one can combine measurements of the proper motions and line-of-sight velocities of an object to estimate the distance based on a rotation model of the Milky Way. This method works exceptionally well for sources on the far side of the Milky Way, where one can exploit the fact that sources on the other side of the Milky Way move in opposite directions, giving them apparent speeds along the longitude of ca.~400~km~s$^{-1}$ or even above. This corresponds to proper motions of several mas per year and is much easier to measure than parallaxes at the mas level. A maser detection with a signal-to-noise of 100 mapped with an angular resolution of 40\,mas would have a positional uncertainty of roughly 0.4\,mas. While for sources within 10~kpc from the Sun, traditional parallax measurements will be more accurate, sources well beyond the Galactic Centre can be robustly and accurately measured using only proper motion measurements. A detailed discussion of Galactic structure based on maser parallaxes with SKA-VLBI can be found in \cite{XuYe01.2026.SKA}.

\section{Evolved stars}
\begin{figure}[!h]
    \centering
    \includegraphics[width=0.6\linewidth]{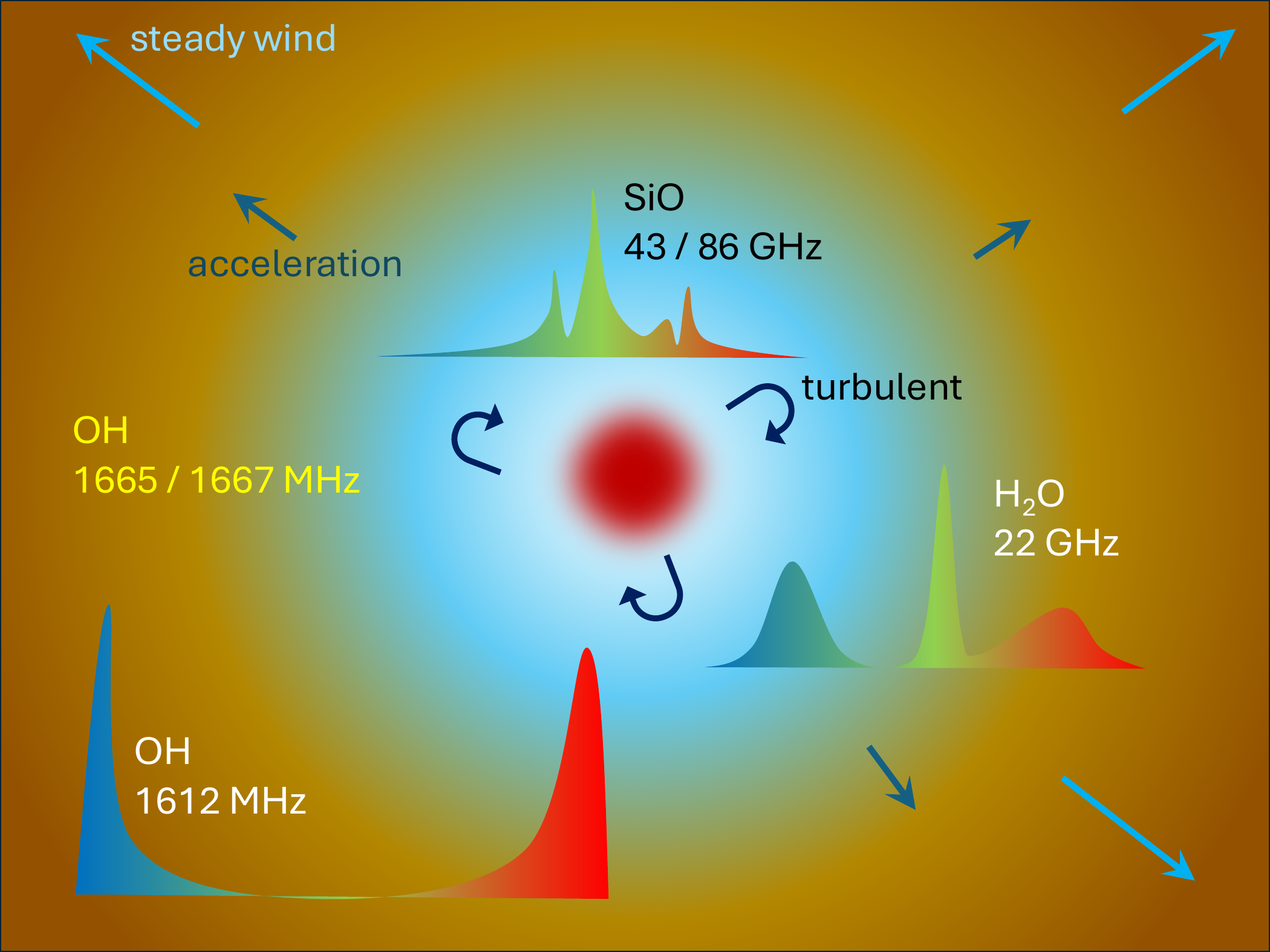}
    \caption{The OH 1612~MHz masers eminently trace the steady wind of dusty evolved stars, causing its double-horned profile. Mainline OH 1665/1667 MHz masers typically trace the more internal parts of circumstellar regions and are susceptible to polarisation due to magnetic fields. H$_2$O masers trace the accelerating part of the wind, whilst the complex and variable SiO masers trace the turbulent extended atmosphere below the dust formation zone. Figure courtesy J.~van Loon.}
    \label{fig:CSEmasers}
\end{figure}

Observing the kinematics of circumstellar matter around an evolved star is a challenging task similar to dense star forming region studies. Also in this case, the natural maser amplification occurs mostly from OH, H$_2$O, SiO masers and allows a view into the gas clouds hidden at most of the wavelengths by dust extinction. The first masers associated with evolved stars were discovered by \citet{wilson1968}: 1612~MHz OH masers were detected in four OH/IR stars of which the red supergiant NML\,Cyg was the brightest. The characteristic double-peaked profile is explained by radial amplification in the steady dust-driven wind of luminous, strongly pulsating cool Asymptotic Giant Branch (AGB) stars and red supergiants (RSGs) \citep{elitzur1976}. 
Today we know that these OH masers originate from the outer circumstellar envelopes (CSEs), many thousand au from the star and allow for detailed kinematic and dynamic studies of gas, while H$_2$O and SiO trace closer regions to the central star (e.g. \citealt{richards2012IAUS}). The 1665/1667-MHz OH mainlines transitions are also commonly found in the CSEs of O-rich evolved stars \citep[e.g.,][]{Lewis:1997, Etoka2004}. While these transitions are generally found tracing more internal regions of the CSEs of the optically-visible (thin-dust shell) Mira stars than the 1612-MHz masers, the 1667 MHz total extent has been found to be greater than the 1612 MHz in (thicker-dust-shell) OH/IR stars \citep{Etoka2010}. Figure~\ref{fig:CSEmasers} shows a sketch of an evolved star with the maser emitting regions. 

The SKA-Mid capabilities will be particularly important in the detection of further targets, especially beyond the Galactic Centre and allowing for the investigation of `Zona Galactica Incognita' as well as in the detection of short-lived phases of stellar evolution as Planetary Nebulae (PNe) in the Galaxy. These capabilities will also provide the combination of high-sensitivity and baseline range needed to obtain high-fidelity imaging of the CSE-sized OH 1612~MHz shell of OH/IR and similar stars throughout the Galaxy and potentially out to the Magellanic Clouds. This has been demonstrated to greatly improve their distance determination via the phase-lag method \citep{Etoka2022}. Distance determination via trigonometric parallaxes, using the ground-state OH maser emission of OH/IR stars, though challenging, has also been proven feasible \citep{Orosz2017}. For a possible applicability of this towards the Galactic Centre see \cite{Imai01.2026.SKA}. Moreover, with SKA-Mid the stellar radio continuum will become detectable for many evolved stars \citep{Bojnordi2024} permitting astrometric registration of various maser species, and a direct comparison of masers with the locations of the ionised jets in PNe.

\subsection{OH masers in the backyard of our Galaxy}
Seen from Earth's vantage point, the stellar population in the backyard of our Galaxy, i.e. the hemisphere beyond the Galactic Centre is not well known. The reason is the general faintness of the stars at distances $>8$\,kpc, the optical
barrier due to dust absorption in the Galactic plane, and crowding of objects, especially on line-of-sights around the Galactic Centre. All
these observational difficulties are strongly mitigated in the case of
stars with luminous radio emission, which are less abundant than
ordinary stars. Such radio beacons are the stars emitting maser
emission from their circumstellar shells, which occurs while low- and
intermediate mass stars ($\approx$1--8\,M$_\odot$) are in the evolutionary
phase on the AGB, or while more massive stars pass the
RSG phase. Especially for OH masers emitting in band 2,
SKA-Mid holds promise for substantial advances.

Currently, close to 3000 OH masers were detected at 1612~MHz toward
evolved stars in our Galaxy \citep{engels2024}. Surveys with 
single-dish radio telescopes discovered most of these masers with flux
density limits of several hundred mJy. The most recent surveys were
however made with interferometers: for example, The HI/OH/Recombination (THOR) line survey of the Milky Way with the VLA \citep{beuther2019}, and the Southern Parkes Large-Area Survey in Hydroxyl (SPLASH) survey with Parkes and Australian Telescope Compact Aarray (ATCA) (\citealt{qiao2020} and references therein), increased the rms sensitivity to $\sim$10 and 70\,mJy beam$^{-1}$  for channel widths of 1.25 and 0.09 km\,s$^{-1}$, respectively. For efficiency reasons, these (blind) surveys, are however restricted to the inner Galactic plane with $\Delta l \pm 70$ deg$^2$ and $\Delta b \pm 2$ deg$^2$.

\begin{figure}[!h]
    \centering
    \includegraphics[width=0.95\linewidth]{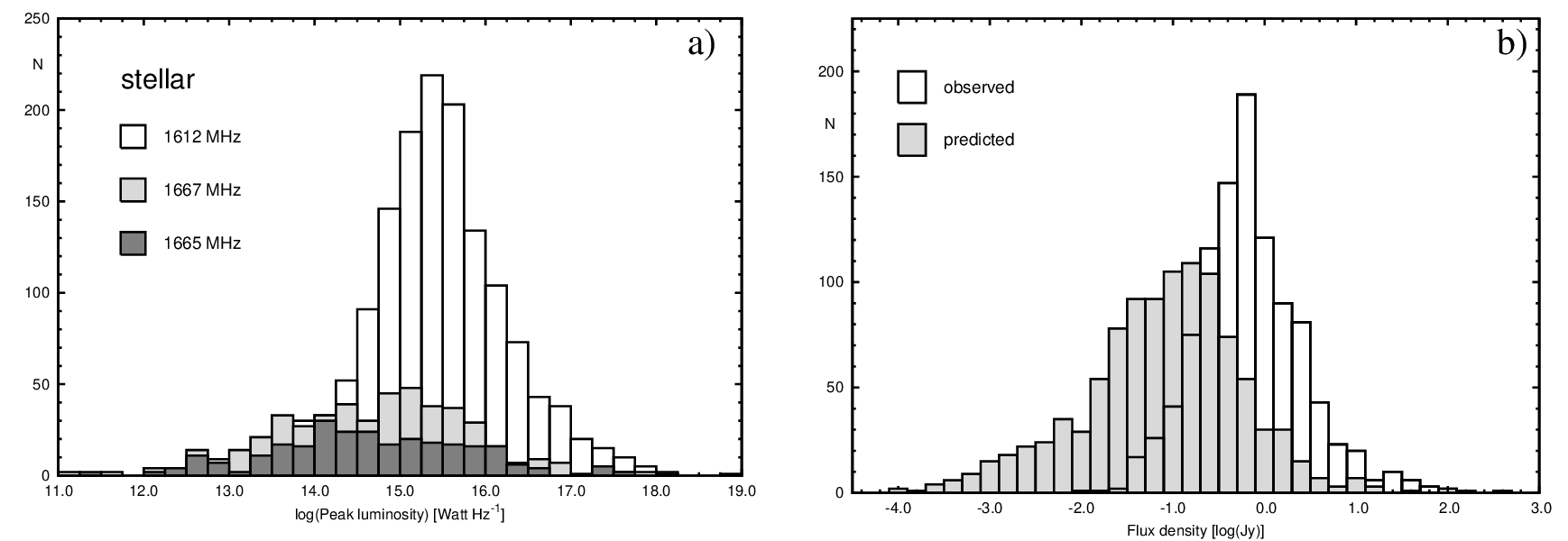}
    \caption{{\blue a)} Distribution of peak luminosities for stellar masers
      assuming kinematic distances. Increasing shading denotes the 1612, 1667,
      1665~MHz transition. {\blue b)} Predicted distribution of 1612-MHz
      stellar OH-maser flux densities. From \cite{etoka2015}.}
    \label{fig:OurBackyard}
\end{figure}

Based on an earlier version of the OH maser catalogue of \citet{engels2024}, \citet{etoka2015} used the OH maser spectra to estimate radial velocities.  For a subsample of the stars located close to the Galactic Plane they then derived kinematic distances, allowing to infer the specific OH maser
luminosity $L_\nu = f_\mathrm{peak} \cdot 4 \pi D^2$, with $f_\mathrm{peak}$ the
peak flux density of the strongest feature in the maser spectrum and
$D$ is the kinematic distance. The distribution of specific luminosities
is shown in Fig.~\ref{fig:OurBackyard}a. There is a steep
increase of the number of stars with decreasing OH maser luminosity
with a maximum at $2 \times 10^{15}$~Watt~Hz$^{-1}$. The decrease in
number at lower luminosities is likely due to incompleteness,
caused by the sensitivity limits of the past surveys. 

The evolved stars with OH maser detections are a sub-group of all
evolved stars, even in the solar neighbourhood, where sensitivity limits
are of lesser importance. Evolved stars can be without OH maser
detection for several reasons. First, the OH
maser observations coverage is incomplete outside the Galactic Plane, where
only pointed observations towards pre-selected samples were
made. Second, in Mira variables, the OH maser brightness varies
typically by a factor of four in response to the stellar pulsations,
and therefore, individual observations, while the star is in its faint
phase, may result in a non-detection. Third, during evolution on the
AGB the luminosity of the OH masers increases with increasing dust
mass-loss rates, leading to low detection rates in the early stages of
AGB evolution. And fourth, the implicit assumption of spherical
symmetry of the OH maser excitation may not be true, as the detection
rate did not pass 60\%, even for stars with appropriate mass-loss
rates and multiple observations at 1612~MHz \citep{lewis1992}; these stars are likely undetected because possible OH maser emission is not beamed into the direction of Earth. 

Using new sensitive variability surveys in the optical and near-infrared, the first samples containing RSGs, semiregular and Mira variables have already been published \citep{albarracin2025, hey2023, zhang2025}. Here, extinction and distances determination may have introduced biases, and radio-selected samples will prove an important contribution. OH maser surveys in particular will allow us to ascertain the presence of O-rich chemistry in the detected stars and pick up the `extreme OH/IR stars', which are strongly obscured in the visual and near-infrared bands.

\citet{etoka2015} predicted the number of OH masers expected in the
opposite hemisphere of our Galaxy. They used a model of the Galactic
distribution of AGB stars and their mass-loss rates and estimated
1612~MHz stellar OH maser flux densities on the base of a relation
between mass loss rates and OH maser luminosities (see Fig.~\ref{fig:OurBackyard}b). They concluded that the bulk of the predicted population has flux densities between 10 and 500 mJy. With SKA-Mid, even in the initial phase, from AA* when the zoom mode becomes
available, a sensitivity of 7\,mJy~beam$^{-1}$ in $\sim$0.6~km~s$^{-1}$ spectral channel can be reached with an integration time of $\sim$2h30m for a pointing towards the Galactic Centre and a uniform weighting hence allowing the detection of this population with a 5$\sigma$ down to a 35~mJy threshold. Band 2 observations with the SKA-Mid will certainly shed new light on the so called `Zona Galactica Incognita': the remote areas of our Galaxy \citep{vallee2014}.

\subsection{OH maser-emitting planetary nebulae}
Planetary nebulae (PNe) are evolved low- and intermediate-mass stars ($<$8 \(M_\odot\)), after having gone through the AGB phase and a brief ($10^2-10^4$ yr) post-AGB phase. OH maser-emitting PNe (hereafter OH-PNe) are an important subset, characterised by presenting both OH masers and radio continuum emission \citep{Zijlstra1989}. As the OH masers are expected to be extinguished shortly after ($\simeq$1000 yr; \citealt{Lewis1989,Gomez1990}) the end of the AGB phase,  the OH-PNe could be in the earliest PNe phases and thus interesting for understanding their formation and evolution. 

To date, only six sources have been confirmed as OH-PNe using interferometric observations 
\citep{Uscanga2012,Qiao2016}. Recently, a few more OH-PNe candidates have been proposed by \citet{Cala2022}. Most of OH-PNe exhibit a bipolar morphology in optical, infrared, and/or radio continuum images, suggesting that the OH maser emission could be related to non-spherical mass-loss phenomena.  OH masers are usually located toward the central region of the nebula \citep{Uscanga2012,Qiao2016}, even tracing an equatorial structure as in IRAS 16333$-$4807 \citep{Qiao2016}. The 1612~MHz OH maser is present in all OH-PNe, while only three 
present the OH maser transition at 1720~MHz \citep{Gomez2009,Qiao2016,Gomez2016}. Interestingly, this transition is only excited by shocks. Moreover, two stars 
present OH maser emission at 6035~MHz \citep{Desmurs2010}, 
which is very rarely found in evolved stars. The SKA-Mid will be important for multi-frequency surveys and studies of PNe.

OH masers can also be used to study the in-situ magnetic fields, e.g., the clear Zeeman splitting observed in the OH maser emission at 1720~MHz in IRAS\,16333$-$4807 \citep{Qiao2016}, which suggests that this transition is formed in a magnetised environment. The typical magnitude for the magnetic fields is of a few mG. Also, OH maser lines in PNe and proto-PNe candidates can exhibit a high degree of circular polarisation $>$\,50\% \citep{Szymczak2004,Gomez2009}.

SKA-Mid will provide the deepest ever unbiased sky survey not only of circumstellar OH masers in AGB stars, but also in very short-lived phases as PNe for which it is critical to increase the number of sources to understand the statistical properties of the entire OH-PNe population and furthermore, to understand the last stages of stellar evolution. Given the improvement in sensitivity by the SKA-Mid array with respect to existing surveys, we expect to find thousands new OH masing sources \citep{dickey2013}. Using SKA-Mid at the initial phase AA* with the zoom mode centred at 1612~MHz, a sensitivity of 2.4 mJy beam$^{-1}$ can be reached for PN\,K\,3$-$35 with an integration time of 1~h for a channel resolution of $\sim$0.08~km~s$^{-1}$, considering robust=0 Briggs weighting. This sensitivity allows us to detect with 5$\sigma$ a 12~mJy threshold enough to identify OH maser lines with a 50\% circular polarisation. Increasing the observing time up to $\sim$3~h, will improve the sensitivity to 1.4~mJy~beam$^{-1}$, allowing us to detect with 5-$\sigma$ a 7~mJy threshold enough to identify OH maser lines with a 30\% of circular polarisation. 

\subsection{Population studies of evolved stars beyond the Galaxy through surveys}

A few aspects of extra-galactic maser surveys in the nearby Universe that will be carried out with the SKA can be found in \href{https://www.skao.int/sites/default/files/documents/d38-ScienceCase_band6_Feb2020.pdf}{Memo 20-01: SKA1 Beyond 15GHz}. With SKA-Mid, we aim to carry out deep searches of OH and
\meth\ masers beyond the Milky Way and beyond the Local Group. In case the frequency coverage of SKA-Mid will be extended beyond 15~GHz, then also the 22~GHz water and 23~GHz ammonia maser transitions will be within reach for our studies of the Local Universe. The latter transitions trace high-density molecular gas in regions hosting peculiar physical conditions, like those found, e.g., in dwarf and starburst galaxies and nearby Active Galactic Nuclei (AGN) \citep[see][for water masers in AGN]{Tarchi01.2026.SKA}. Here, we describe the opportunities with SKA-Mid for studies of dust-driven winds in evolved stars at low metallicity using g-OH masers.
 
While the OH luminosity is related to the mass-loss rate of OH/IR stars (the maser is pumped through OH transitions in the mid-IR), it does not provide a measure of the gas mass-loss rate that is independent of the SED-derived dust mass-loss rate \citep{zijlstra1996}. The main power in observing 1612~MHz OH masers lies in the fact that the wind speed can be measured; the narrowing of the maser lines and intrinsically high spectral resolving power at radio frequencies allows accurate measurements of typical wind speeds in the range 5--30\,km\,s$^{-1}$ (\citealt{sevenster2001} and references therein). Not only is this needed to scale the SED-derived optical depth to a mass-loss rate, but it also offers a means to derive the dust-to-gas ratio, assuming the wind is driven by radiation pressure on dust or, alternatively, a means to test this assumption.
The radial amplification and non-linear nature of the maser mechanism often result in highly asymmetric 1612~MHz OH maser profiles, and sometimes substructure in the peaks (e.g. \citealp{marshall2004}). Especially at low signal-to-noise this can introduce uncertainties in wind speed measurements. 1665/7~MHz OH maser detection can mitigate this, because they arise from the extended atmospheric layers that lie deeper in the circumstellar shell/envelope where the wind is accelerated/still in acceleration. 
The 1665/7~MHz OH maser profile therefore, often shows peaks of emission close to the stellar velocity, providing a reference for interpreting the 1612~MHz OH maser profile. Water and SiO masers can also be used for this, but they fall outside the present SKA-Mid bands. 

Extragalactic maser populations have the advantage of relatively well-known distances, especially so among the population itself, and therefore allow better determined luminosities and mass-loss rates. They also trace a larger variety of environments than what is found within our Galaxy. For instance, OH/IR stars in the Large Magellanic Cloud (LMC) and the Small Magellanic Cloud (SMC) at distances around 50--60\,kpc have metallicities $Z_\mathrm{LMC}\approx 0.5\ Z_\odot$ and $Z_\mathrm{SMC}\approx 0.2\ Z_\odot$. Galactic populations at such low metallicities are limited to low-mass stars found in the thick disk, halo and bulge, but in the LMC and SMC, also massive AGB stars and RSGs are present at such metallicities.

The first extragalactic evolved star maser discovered was the 1612~MHz and 1665~MHz OH emission from the extreme red supergiant WOH\,G64 in the LMC \citep{wood1986}, followed by five further detections in the LMC, including four massive AGB stars \citep{wood1992}. The 1612~MHz OH maser profile of WOH\,G64 proved to be one of those deceptive cases where sub-structure in the blue peak was interpreted as a very slow wind, whereas subsequent detections of SiO maser emission \citep{vanloon1996} and H$_2$O maser emission \citep{vanloon1998b} revealed a kinematic offset, revising the wind speed upwards. Eventually, the expected red peak was detected in 1612~MHz OH maser emission \citep{marshall2004}.

OH maser observations in the LMC were mostly conducted with the Parkes single-dish telescope at first, and later using the ATCA interferometer. After \citet{vanloon1998a}, \citet{marshall2004}, and \citet{goldman2017} refined the target samples to increase the chances of success, the tally now stands at 15. This OH/IR population revealed clear evidence for the expected scaling of wind speed with luminosity and dust-to-gas ratio. No OH/IR stars have been found in the SMC  \citep{goldman2018}, almost certainly implying less efficient masing. Both the LMC and SMC searches reached down to rms noise levels of 4\,mJy, and with SKA-Mid these could be pushed down further. If so, there would be a high chance of detecting the first evolved-star OH maser in the SMC, which would provide a stringent new test on dust-driven winds at low metallicity.

The nearby galaxy NGC\,6822 may provide another opportunity to search for evolved star OH masers at a metallicity below that of the LMC. It, too, boasts a sizeable population of OH/IR star candidates, including red supergiants. However, at a distance of 0.5\,Mpc it would require $<0.1$\,mJy noise levels to be comparable to the deepest SMC observations, or to have some rare, exceptionally bright sources to be present. It may also be possible to detect the brightest 1612~MHz OH masers of RSGs in M\,33 at a distance of 0.8--0.9\,Mpc, if sub-mJy rms noise levels can be reached. Indeed, a tentative discovery of a maser counterpart was presented by \citet{kinson2022}. 

\section{Masers for science verification and calibration of maser data}

Masers, as bright, compact sources with a narrow spectral line profile can play a role in the early testing of the SKA-Mid array. Although OH masers are intrinsically variable, several well-known high-mass star-forming regions show persistent spectral structure and strong polarised components over decades. Examples suitable for SKA-Mid science verification at 1665/7 MHz include W49N \citep[$\sim$250 Jy,][]{Bayandina2021} and G351.417+0.645 \citep[$\sim$400 Jy,][]{Caswell2013}. These sources are bright, compact, accessible from the Southern hemisphere, and have extensive variability histories as well as ongoing single-dish monitoring. In addition, there are many bright ($>$ 100\,Jy) and compact methanol masers. Finally, sensitive simultaneous SKA-Mid observations of multiple maser transitions will need to handle a very wide dynamic range and various spectral resolutions, making it important to keep calibrated target visibilities. 
The high SNR of maser spots can be used for self calibration of the target visibility data to improve image quality.

\section{Conclusions}

Maser emission remains one of the most powerful and versatile tracers of physical conditions in the Universe. Thanks to their compactness and brightness masers provide precise astrometry, with which astronomers can study environments spanning a wide range of astrophysical scales and stellar evolutionary stages — from accretion disks around the youngest massive stars and the envelopes of evolved low-mass stars in the Milky Way to the Galactic disk structure and evolved stars in nearby galaxies. In all these cases, 
the remarkably narrow maser lines provide extraordinarily detailed information: they reveal the kinematics and structure of the gas, constrain temperatures and densities, trace magnetic fields, and yield distances and evolutionary stages of the target sources. The upcoming capabilities of SKA-Mid promise to transform this field. With its unprecedented sensitivity and broad instantaneous frequency coverage, the SKA will uncover new, fainter maser populations and enable studies of previously unexplored 
regions within our Galaxy and beyond. These observations will not only expand our maser catalogues but also refine our understanding of the processes that drive star formation, stellar evolution, and Galactic dynamics. 

\section{Acknowledgements}
The authors are very grateful to the anonymous referee for their detailed and constructive comments. A.B. acknowledges support from the National Science Centre, Poland through grant 2021/43/B/ST9/02008. A.M.J.  would like to acknowledge the support of the Max Planck Society and the Collaborative Research Center 1601 (SFB 1601 sub-project A1 and B2) funded by the Deutsche Forschungsgemeinschaft (DFG, German Research Foundation) – 500700252.

\bibliographystyle{abbrvnat-maxbibnames4}
\bibliography{librarian} 

\end{document}